\documentclass[useAMS,usenatbib,usegraphicx]{mn2e}

\title[Turn-off of deuterium astration in the Galaxy disk]{Turn-off of deuterium astration in the recent star formation of the Galaxy disk}
\author[T. Tsujimoto]{T. Tsujimoto$^{1}$\thanks{E-mail:
taku.tsujimoto@nao.ac.jp} \\
$^{1}$National Astronomical Observatory of Japan, Mitaka, Tokyo 181-8588, Japan\\}

\def\ltsima{$\; \buildrel < \over \sim\;$}
\def\ltsim{\lower.5ex\hbox{\ltsima}}
\def\gtsima{$\; \buildrel > \over\sim \;$}
\def\gtsim{\lower.5ex\hbox{\gtsima}}
\def\ms{$M_{\odot}$ }
\def\msp{$M_{\odot}$}

\begin{document}

\date{Accepted 2010 September 3. Received 2010 September 2; in original form 2010 July 26}

\pagerange{\pageref{firstpage}--\pageref{lastpage}} \pubyear{2010}

\maketitle

\label{firstpage}

\begin{abstract}
Chemical features of the local stellar disk have firmly established that long-term, continuous star formation has been accompanied by a steady rate of accretion of low-metallicity gas from the halo. We now argue that the recent discovery of an enhanced deuterium (D) fraction in the Galaxy is consistent with this picture. We consider two processes: the destruction of D in the interior of stars (astration) and the supply of nearly primordial D associated with the gas infall. Conventional Galactic chemical evolution models predict a monotonic decrease in D/H with time with a present-day D/H abundance which is much lower than the local value recently revealed. This predicted feature is the result of high levels of deuterium astration involved in the formation of the local metal-enhanced disk. Here we propose a new channel to explain the observed enhancement in D/H. Our model, which invokes ongoing gaseous infall and a star formation rate that declines over the past several Gyr, predicts that the D astration is suppressed over the same time interval.
\end{abstract}


\section{Introduction}

During the past decade, there has been a steady improvement in observations of deuterium to hydrogen (D/H) in the universe. The measurement accuracy is now good enough to provide insight into Galaxy evolution. First of all, the primordial D/H ratio (D/H)$_p$ is confined to a narrow range of $\sim$(2.6-2.8)$\times 10^{-5}$. This range covers the analysis result of data from the {\it Wilkinson Microwave Anisotropy Probe} (WMAP)  that gives 2.60$^{+0.19}_{-0.17}$$\times 10^{-5}$ \citep{Coc_04} or 2.75$^{+0.24}_{-0.19}$$\times 10^{-5}$ \citep{Cyburt_03}, depending on the different light-element nuclear reaction rate \citep[also 2.57$\pm$0.15 from the WMAP 5 data;][]{Dunkley_09}, and the measurements of distant metal-poor damped Ly$\alpha$ systems that lead to 2.82$^{+0.20}_{-0.19}$$\times 10^{-5}$\citep[][see also O'Meara et al.~2006]{Pettini_08}. Recent {\it Far-Ultraviolet Spectroscopic Explorer} (FUSE)  observations reveal that the present-day D/H abundance (D/H)$_0$ for interstellar matter (ISM) in the Galactic disk is  surprisingly as high as $\geq$ 2.31$\pm$ 0.24$\times 10^{-5}$ after correcting for dust depletion \citep{Linsky_06}. Some authors claim there are residual effects due to dust depletion which can account for the large scatter in the observed D/H  \citep{Steigman_07, Ellison_07}. But the high D/H abundance of 2.2$^{+0.8}_{-0.6}$$\times 10^{-5}$ in the warm neutral medium (WNM) of the lower Galactic halo, which originates in the disk and is elevated into the halo \citep{Savage_07}, strongly supports the claim by \citet{Linsky_06}. 

The difference between (D/H)$_p$ and (D/H)$_0$ is defined as a deuterium astration factor $f_d$=(D/H)$_p$/(D/H)$_0$. The value of $f_d$ should exceed unity because it is believed that all D is produced during Big Bang nucleosynthesis, and any subsequent destruction process of D occurs in the interior of stars. Thus, the value of $f_d$ is determined by the fraction of matter which has never been cycled through stars in the present ISM. Accordingly, $f_d$ is closely related to the degree of heavy-element enrichment through star formation. Our expectation is that the D abundance in the ISM is anti-correlated with the metallicity such that the well-enriched ($\sim$solar metallicity) local disk demands a high $f_d$. Indeed, conventional Galactic chemical evolution (GCE) models have been applied to the evolution of D so far by several authors \citep{Prantzos_96, Tosi_98, Matteucci_99}, and predict $f_d=1.6\pm 0.2$ for the local disk \citep{Romano_06}. In other words, the local abundances of heavy elements such as Fe or $\alpha$-elements in the ISM as well as long-lived stars imply $f_d \approx 1.4$ at the least, that is equivalent to  (D/H)$_0 \ltsim $2.0$\times10^{-5}$. On the other hand, the high (D/H)$_0$ suggested by the recent observations implies a smaller $f_d$. For the values of (D/H)$_p$=2.75$\times 10^{-5}$ and (D/H)$_0$=2.31$\times 10^{-5}$, we obtain $f_d$=1.19, which is incompatible with the theoretical prediction by GCE models.

The D/H abundances which bridge between the primordial (D/H)$_p$ and the present-day (D/H)$_0$ provide crucial information on the evolution of D. The D/H abundance in the protosolar cloud provides crucial information at a look-back time of $\sim$ 4.6 Gyr. The D/H abundances deduced from the Jovian atmosphere \citep{Lellouch_01} and from the solar wind \citep{Geiss_98} are estimated to be $2.1\pm$ 0.4 (0.5)$\times 10^{-5}$.  This result is broadly consistent with the present-day local value. If these values are representative of the local disk, we are led to believe that  D astration involved with star formation has turned off for the last several Gyr.

In fact, as the process to change the amount of D in the ISM, the Galactic disk involves an enrichment process through an infall of low-metallicity gas from the halo, which brings (nearly) primordial D, besides its destruction through star formation. This infall is required by the GCE models to reproduce the abundance distribution function (ADF) of local disk stars \citep[e.g.,][]{Pagel_97}.  In addition, we cite high-velocity clouds (HVCs) as the evidence for the present-day infall \citep[e.g.,][]{Wakker_99, Lockman_08}. The infall operates in the GCE models so as to enrich the ISM efficiently in the early phase to suppress the number of metal-poor stars, and maintain the star formation activity by continuously suppling the gas. This established view predicts a monotonic increase (decrease) in heavy elements (D) in the ISM up to the solar abundances ((D/H)$_0 \ltsim$2.0$\times10^{-5}$). In this way, the evolution of D is tightly bound to the production of metals in star formation. 

But what if a significant fraction of the heavy elements in the disk does not originate from {\it in situ} star formation \citep[e.g.][]{Oppenheimer_10}? This represents a radical departure from conventional GCE models and has recently been discussed in detail. Here large-scale winds from the Galactic bulge, which entrain a large amount of heavy elements, enrich the disk. The hypothesis well explains the origin of metal-rich stars in the solar neighborhood \citep{Tsujimoto_07} and the mechanism of the time-evolution of abundance gradient \citep{Tsujimoto_10}.  In this scheme, a part of heavy elements trapped in long-lived stars as well as in the gas of the disk is not related to the local star formation but the activity of the bulge.

In this paper, we shed light on the evolution of deuterium in the framework that the Galaxy disk has evolved through "wind + infall", together with the renewed constraint from the star formation history in the local disk. Then we show that the combined phenomena of winds that entrain heavy elements and infall that contains primordial D will make the Galaxy disk feature withstand the test of observations.

\section{Wind Scenario}
In this section, we summarize the proposed wind scenario, which is a key ingredient in the making of a heavy-element feature in the disk, based on the discussions in \citet{Tsujimoto_07} and \citet{Tsujimoto_10}, together with the updated observational results.

\subsection{Signature of wind enrichment on the Galaxy disk}

\noindent {\sl 1. Metal-rich stars in the solar neighborhood}
 
There exist metal-rich stars beyond the solar metallicity, the fraction of which is roughly 20\%. In fact, its presence as well as its elemental feature poses a puzzling problem. The ADF of solar neighborhood disk stars has been studied by many authors \citep[e.g.,][]{Wyse_95,Rocha-Pinto_96,Nordstrom_04}, and the location of its peak at [Fe/H]$_{\rm peak}$$\approx$ -0.15 has been firmly established. On the other hand, the metal-rich end of the ADF extends at least to [Fe/H]$\sim$ +0.2. Spectroscopic observations of elemental abundances for metal-rich disk stars \citep{Feltzing_98,Bensby_05} have confirmed that chemical enrichment in the solar neighborhood continued unabated until [Fe/H]$\sim$ +0.4. \citet{Tsujimoto_07} has claimed that the simultaneous reproduction of both the presence of stars with [Fe/H]\gtsim +0.2 and [Fe/H]$_{\rm peak} <$0 is hard to realize through the conventional scheme of local enrichment via low-metallicity gaseous infall from the halo, and that the presence of supersolar stars is crucial evidence for enrichment by winds from the bulge. It should be of note that the seeming making of supersolar stars by GCE models thus far \citep[e.g.,][]{Yoshii_96, Yin_09, Kobayashi_09} is an end result of convolution of the ADF with a dispersion of 0.1 - 0.2 dex in [Fe/H].

The convincing signature of wind enrichment is seen in their elemental features. Nearby metal-rich stars are represented by an upturn in [$\alpha$/Fe] ratios except for [O/Fe] against increasing [Fe/H]\citep{Reddy_03, Bensby_05}. Such upturns can also be seen in other elements, such as odd elements (Na, Al), and some iron-group elements (Ni, Zn). It is incorrect to attribute these upturns to metallicity-dependent SN II yields, especially because this feature is observed with both even- and odd-numberd elements. These features can be interpreted as a reflection of the elemental feature of winds from the bulge \citep{Tsujimoto_07}. Since the [$\alpha$/Fe] ratios start to upturn from [Fe/H]$\sim 0$, the wind enrichment is expected to occur around 5 Gyr ago, corresponding to the time of sun's birth. From the above discussion on elemental feature, the galactic fountain \citep[e.g.,][]{Shapiro_76, Bregman_80} that may serve as a means for transporting metals to other locations in the disk is unlikely to contribute to acceleration of the chemical enrichment in the solar vicinity.
 
On the other hand, radial mixing of disk stars due to resonant scattering with transient spiral waves predicts that contaminants coming from the inner disk populate the solar neighborhood (Sellwood \& Binney 2002; Ro\v{s}kar et al.~2008a,b; Sch\"{o}enrich \& Binney 2009; S\'{a}nchez-Bl\'{a}zquez et al.~2009), and allows the presence of metal-rich stars beyond the upper limit of a local chemical enrichment \citep{Roskar_08b}. However, stars orbiting near the Galactic center should exhibit a decreasing [$\alpha$/Fe] trend with increasing [Fe/H], as observed in the bulge \citep{Fulbright_07, Melendez_08}. Therefore, elemental features of nearby metal-rich stars seem at odds with the prediction of radial mixing as a mechanism of producing metal-rich stars. It should be of note that some authors claimed no upturning feature but a constant solar ratio for nearby metal-rich stars \citep{Chen_08}. However, none of the existing models including radial mixing supports this result.
 
\noindent {\sl 2. Time evolution of the abundance gradient}

\citet{Tsujimoto_10} has found that the radial [Fe/H] gradient roughly from $R_{\rm GC}$=4 to $R_{\rm GC}$=14 kpc ($R_{\rm GC}$: Galactocentric distance) has flattened out in the last several Gyr with a change in a slope of $\sim $-0.1 dex kpc$^{-1}$ to $\sim$ -0.05 dex kpc$^{-1}$, by comparing the steep relic gradient observed in the open clusters with that observed in the Cepheids \citep[see also][]{Chen_03, Maciel_06}.  Chemical evolution models predict contradictory time evolution of the metallicity gradient, i.e., a steepening \citep{Chiappini_01} or a flattening \citep{Boissier_99, Hou_00}.  Putting aside these contradictions, existing GCE models predict a monotonic increase in abundances for each region, and the predicted gradient change is small compared with the observations. \citet{Tsujimoto_10} has claimed that large-scale winds from the Galactic bulge are the crucial factor for its mechanism. Their proposed scenario is (i) winds once set up a steep abundance gradient ($\sim$ -0.1 dex kpc$^{-1}$) owing to the enrichment by heavy elements entrained in the winds, and (ii) later evolution leads to a flattening of abundance gradient through chemical evolution under an accretion of a low-metal gas from the halo.

On the other hand, new analysis of planetary nebulae supports a slight steepening of the [O/H] gradient with time \citep{Stanghellini_10}. This contradictory result is interesting because it is likely that winds from the bulge are not so metal-rich in O. The metal-rich bulge stars exhibit the low [O/Fe] below the solar ratio \citep{Fulbright_07}. Abundance of the microlensed stars recently analyzed confirmed [O/Fe]$\sim$ -0.2 for metal-rich bulge stars \citep{Bensby_10}. The reduced O yield of SN II expected from these observations is probably due to efficient mass loss for massive stars, which prevents He and C from synthesizing into O within the metal-rich environment in the bulge \citep{McWilliam_08}. Thus, O is not a good indicator for the wind enrichment that is predicted to cause the time evolution of the abundance gradient in our scheme. In other words, the time evolution of the [O/H] gradient could be possibly discussed in the conventional GCE scheme as predicted by \citet{Chiappini_01}. Alternatively, it is possible that some effect by radial mixing which causes a steepening of abundance gradient \citep{Roskar_08b, Sanchez_09} is seen in the evolution of O gradient owing to little influence of the wind enrichment.

In addition, \citet{Stanghellini_10} compiled a new data set of ages, [Fe/H], and distances for open clusters. From their samples, we make two groups: one is composed of young clusters whose ages are less than 1 Gyr, and the other is of old ones whose ages range within $5\pm2$ Gyr. Here we restrict both of samples to the ones with $R_{\rm GC}<$14 kpc. As a result, the least-square fitting gives a slope of -0.05 dex kpc$^{-1}$ for young clusters, and  -0.14 dex kpc$^{-1}$ 
for old clusters, respectively. We see a significant flattening of the [Fe/H] gradient for new open cluster data.

\noindent {\sl 3. Other signatures}

In the Galaxy, there is now strong evidence for large-scale winds across the electromagnetic spectrum \citep{Bland_03, Fox_05, Keeney_06, Everett_10}. Furthermore, there is tantalizing direct evidence that dust and gas from the inner Galaxy have been transported to the solar neighbourhood \citep{Clayton_97}. In standard chemical evolution models, the isotopes $^{29}$Si and $^{30}$Si become more abundant than $^{28}$Si as the disk ages. But pre-solar grains from meteorites show evidence that the local isotopes formed in material that had experienced more nuclear synthesis than material that came 
after the Sun's birth. 

\subsection{Driver of winds}

A driver of large-scale winds would be a starburst in the Galactic bulge. Therefore, our scenario proposes that the Galactic bulge has a complex history involving at least one starburst episode around 5 Gyr ago in addition to the initial burst that gave rise to the old population $>$ 10 Gyr ago. To date, the star formation history of the Galactic bulge still remains an open question because we do not have the precise color-magnitude diagram (CMD) of the bulge owing to the difficulty of discrimination of bulge stars from disk stars \citep[e.g.,][]{Holtzman_93, Feltzing_00}. There is good evidence for recent star formation in the nuclear region \citep{Launhardt_02}. In addition, from the perspective of bar formation, one might expect some star formation activity in the region of what is now the bulge, about 3 Gyr after the inner disk started to form - this would be at the time when the bar buckling occurs \citep{Athanassoula_09} - it would really stir up any remaining gas in the inner disk.

It is difficult to conjecture how massive a starburst is likely to be from a viewpoint of a wind enrichment on the disk since there are too many uncertain factors to predict it. Here we take a simplified case and show that a starburst which produce $\sim 5\%$ of the bulge stars might be one candidate, by checking supply and demand of Fe. For simplicity, the disk is here considered to be one zone represented by the solar circle. Let us take the masses  of a bulge and a disk and the gas fraction of a disk as $2\times 10^{10}$\msp, $4\times 10^{10}$\msp, and 0.1. The IMF with a slope of $x$= -1.05 \citep{Matteucci_90, Ballero_07} together with the star formation producing the fraction of $5\%$ of bulge stars yields $\sim 3\times 10^7$ SNe II that eject $\sim 4\times 10^6$ \ms of Fe, some fraction of which will be, though, retained within the bulge.  On the other hand, the Fe abundance of gas in the disk should increase from [Fe/H]=0 to [Fe/H]=+0.3 by the wind enrichment plus the local enrichment by star fromation. It means that the winds should entrain roughly the same amount of Fe as that already present in the ISM to double the Fe abundance if we neglect the local Fe production.  In the end, the total Fe mass necessary for the additional enrichment on the disk amounts to $4\times 10^{10}\times 0.1 \times 10^{-3} ({\rm solar \ Fe \ abundance}) \sim 4\times 10^6$\msp, equivalent to the former estimate. If we assume that the wind enrichment is restricted to the inner disk including the solar vicinity together with a partial contribution to the winds from type Ia supernovae (SNe Ia) and hypernovae that eject high Fe masses, the necessary fraction of stars produced in a burst will significantly decrease like  $<$ 1\%. 

While numerous studies \citep{Ortolani_95, Feltzing_00, Kuijken_02, Zoccali_03, Clarkson_08} agree that the bulk of the bulge formed $>$10 Gyr, a minority population of stars as young as 5 Gyr cannot be ruled out, and a minor burst of star formation as recent as 5 Gyr may have taken place in the bulge. The fraction of stars produced by such a minor burst must be small like $\sim 5$\% as implied by an above simplified estimate so as not to contradict the present CMDs, but in practice, how much of stars  the starburst was likely to produce remains quite unclear. As well, the age of a burst claimed here is not definitive due to a lack of determinant factor of its age-dating. The results from microlensed dwarf and subgiant stars in the Galactic bulge showing the average age of $\sim$ 7 Gyr for metal-rich ([Fe/H]$>$ 0) bulge stars \citep{Bensby_10} may be in favor of our hypothesis.

Recently, some authors claim that momentum-driven winds are an efficient mechanism for a powerful wind from the central starburst \citep{Murray_05, Oppenheimer_06, Oppenheimer_10}. These strong winds could realize a widespread enrichment over the disk, possibly including the outer disk region, through  hydrodynamical interaction between the winds and the gaseous halo \citep{Bekki_09, Oppenheimer_10}. In N-body hydrodynamic simulations,  only high-resolution models can create inhomogeneous ISM sufficient for the right treatment of winds from star-forming region \citep[e.g.,][]{Ceverino_09}. Ongoing advanced numerical calculations will certainly reveal a full view of the wind enrichment.

\subsection{Influence of an initial starburst in the Galactic bulge}

The main body of the Galactic bulge was considered to be formed in a starburst $>$ 10 Gyr ago. It is natural that we expect the action of ravaging winds associated with this vast star formation. Since the age of thin disk is as low as $\ltsim$ 10 Gyr \citep[e.g.,][]{Hansen_02, Peloso_05}, the signature of initial wind enrichment is likely to be unseen in the thin disk stars. Instead, some fossil record should be imprinted in the formation of thick disk prior to thin disk. Indeed, recent study claims that ages of the thick disk stars are $>$8-9 Gyr with a span of $\sim$3 Gyr \citep{Bensby_07}, which suggests a similar formation epoch between thick disk and bulge. Here we discuss compelling  evidence for the contamination of a gas used for the thick disk formation by heavy elements entrained in the winds.

The ADF of thick disk shows a sharp increase from [Fe/H]$\sim -1.3$ to the peak located at [Fe/H]$\sim -0.7$. This feature is first reported by \citet{Wyse_95} with a small number of but complete samples, and the location of its peak is confirmed by a huge SDSS data base of thick disk stars  \citep{Allende_06}. These observational results suggest that a material for the proto-thick disk 
was pre-enriched up to [Fe/H]$\sim$ -1.3 at $>$ 10 Gyr ago. Our claim is that the mechanism of  pre-enrichment is attributable to the wind occurrence triggered by an initial starburst in the Galactic bulge. Detailed results of not only the ADF but also the evolution of [$\alpha$/Fe] calculated with the wind models will be presented in Tsujimoto (2010, in preparation). In addition, similarity of stellar elemental features between bulge and thick disk recently revealed \citep{Melendez_08, Bensby_10} seems to have some connections to the wind phenomena during their overlapping formation episode.

From this viewpoint, metal-weak thick disk stars can be discussed. \citet{Chiba_00} found that some fraction of low-metallicity stars in the range of $-2.2\leq$[Fe/H]$\leq -1$ belong to the thick disk population. According to \citet{Caimmi_08}, addition of these stars to the ADF deduced by \citet{Wyse_95} does not form a metal-poor tail. Rather, it produces another distinct distribution for [Fe/H]$<-1.4$. In our view, metal-weak thick disk stars are regarded as the stars that are formed prior to the significant influence of metal-rich winds from the bulge on the proto-thick disk.

\section{Model and Observational Constraints}

\subsection{"wind+infall" model}

We incorporate the enrichment process by large-scale winds into the standard GCE model. The basic picture is that the disk was formed through a continuous low-metal infall of material from outside the disk region, and for a specific period heavy elements carried by winds from the bulge dropped and enriched the disk. The period of the occurrence of winds is set to be 4-6 Gyr ago so as to reproduce the chemical features of metal-rich solar neighborhood stars as well as the observed flattening of abundance gradient in the last several Gyr \citep{Tsujimoto_10}. The abundances of the winds are set to be identical to the ejecta of type II SNe (SNe II) such as [Fe/H]=+0.4, [Mg/Fe]=+0.4, and [O/Fe]=-0.2. A low O abundance in the wind  is implied by the bulge abundance \citep{Fulbright_07, Tsujimoto_07, Bensby_10}. Details of the models including the equations to be solved are described in \citet{Tsujimoto_10}. The key model ingredients, infall rate and star formation rate (SFR), are presented in the following separate subsections. On other model inputs, their individual descriptions are simply mentioned below.

The initial mass function (IMF) should be chosen carefully in this study since it determines the returned fraction $R$, i.e., the fraction of stellar mass that is returned to the ISM, which crucially influences the evolution of D \citep[e.g.,][]{Prodanovic_08}. We adopt a power-law mass spectrum with a slope of -1.35, together with the mass range [0.05 \msp, 50 \msp]. This combination is deduced from the theoretical analysis considering two conditions given by the chemical abundances of halo stars and the mass-to-luminosity ratio in the solar neighborhood \citep{Tsujimoto_97}. Then, this adopted IMF is combined with the nucleosynthesis yields of SNe II and type Ia SNe (SNe Ia) taken from \citet{Tsujimoto_95}. For the occurrence frequency of SNe Ia, we assume that the fraction of the stars that eventually produce SNe Ia for $3-8$\ms is 0.05 with the lifetime of SN Ia progenitors of 0.5$-$3  Gyr \citep{Yoshii_96}. Note that the returned fraction appears in the equations to solve the evolutions of gas and abundances (O, Mg, Fe) with no instantaneous recycling approximation.

The evolution of D is described by a simple equation that is governed by the time changes of a gas fraction $f_g$, a SFR $\psi(t)$, and an infall rate $A(t)$;

\begin{equation}
\frac{d(Df_g)}{dt}=- D\psi(t) + D_pA(t)  \ \ ,
\end{equation}

\noindent where D in the winds is neglected since the winds mostly consists of the ejecta of SNe II with no D. The D abundance in the infall is assumed to be primordial since the metallicity ([Fe/H]) of the infalling gas is expected to be much lower than the present-day infall such as Complex C whose metallicity is subsolar and D/H is estimated to be 2.2$\pm$0.7$\times 10^{-5}$ \citep{Sembach_04}. In the following, we will describe the key ingredients that determine the D/H evolution, together with the observational constraints to each of them.

\begin{figure}
\includegraphics[width=230pt]{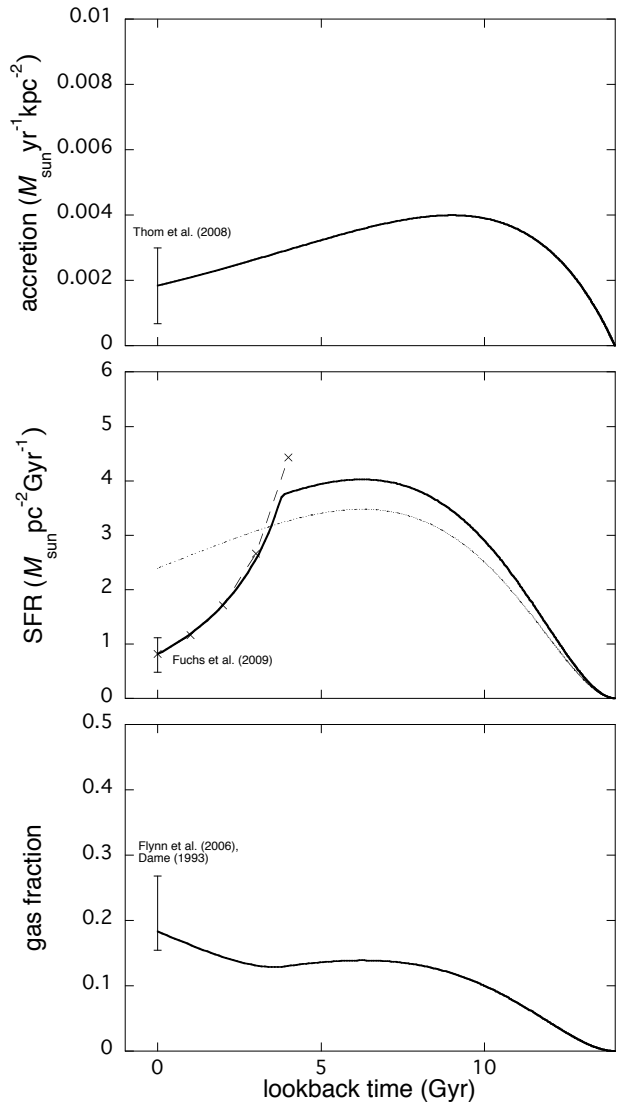}
\caption{{\it upper panel}: The model infall rate as a function of lookback time, together with the observed present-day rate which is estimated from Complex C \citep[][; see, however, the discussion in the text]{Thom_08}. The model infall rate is normalized so that its integration over the whole age gives a total surface density of 42 \ms pc$^{-2}$. {\it middle panel}: The calculated star formation histories. Two model cases are considered. One (solid line) involves a setup of a decreasing SFR for 0$< t<$4 Gyr so as to be compatible with the observed tendency (dashed line) derived by \citet{Fuchs_09}. The other (dotted line) is the one with an assumption of a constant SFR coeficient. Both models are normalized to become the present-day stellar surface density of 35.5 \ms pc$^{-2}$. The observed SFR at the present time is deduced from the mass growth of the stellar disk in the last 1 Gyr \citep[see][]{Fuchs_09}. {\it lower panel}: The evolution of a gas fraction calculated with the decreasing SFR model, together with the present-day value.}
\end{figure}

\subsection{infall rate}

For the infall rate, we adopt the formula that is proportional to $t\exp(-t/t_{\rm in})$ with a timescale of infall $\tau_{\rm in}$. The value of $\tau_{\rm in}$ is assumed to be 5 Gyr so as to reproduce the observed feature of metal-poor tail of the ADF \citep{Yoshii_96}. To compare with the observations, we give an infall rate in units of \ms yr$^{-1}$kpc$^{-2}$, that is shown on the upper panel of  Figure 1. Here we normalize this function so that its integration over 14 Gyr sums up to today's total surface density in the local disk. Today's density is estimated to be 42 \ms pc$^{-2}$ from the surface density of the stellar disk \citep[35.5 \ms pc$^{-2}$;][]{Gilmore_95, Flynn_06} and of the interstellar gas \citep[6.5 \ms pc$^{-2}$;][]{Dame_93}. This together with our infall function predicts the present rate of 0.00185 yr$^{-1}$kpc$^{-2}$. On the other hand, Complex C gives an associated mass flow of 0.00067 - 0.003 yr$^{-1}$kpc$^{-2}$ \citep{Thom_08}. In the figure, we adopt these values as an observationally implied range. However, the average rate given by the entire HVC system may be higher by a factor of $\sim 2$ as predicted so far \citep[e.g.,][]{Chiappini_01, Peek_08}. Still, our prediction for the present-day rate falls within the observationally possible  range. Furthermore, it should be of note that a significant number of ionized high-velocity gas detected in the vicinity of the Galaxy must be a factor to increase the present-day rate \citep[e.g.,][]{Sembach_03, Fox_04}.

\subsection{star formation rate}

The SFR is assumed to be proportional to the gas fraction $f_g$ with a coefficient $\nu$ (Gyr$^{-1}$). In the standard GCE models, the value of $\nu$ is set to be constant over the age of a galaxy. On the other hand, recent several works reveal the star formation history in the local disk especially for the last several Gyr \citep{Hernandez_00, Vergely_02, de la Fuente Marcos_04, Cignoni_06, Fuchs_09}. The common feature claimed by them is that the SFR has been clearly declining with time for the last several Gyr. For the purpose of its reproduction, we introduce  the time-dependent $\nu$ for the recent SFR so as to obey the observed SFR-$t$ relation. Using the relation found by \citet{Fuchs_09} giving a decreasing rate of $\sim$ -0.8 \ms Gyr$^{-1}$, we assume $\nu$=0.6 for $t\leq 10$ Gyr and $\nu$=0.6$\times \exp(-0.5\times (t-10))$ for 10$< t\leq $ 14 Gyr. The middle panel of Figure 1 shows the history of SFR where its normalization is done so that the present-day stellar density beomes 35.5 \ms pc$^{-2}$. For reference, the case with a constant $\nu$ (=0.6) over the full age is indicated by dotted line, demonstrating that a constant $\nu$ model cannot reproduce the observed SFR over the last Gyr.

The present SFR is deduced from the mass growth of stellar density over the last 1 Gyr. From the result of \citet{Fuchs_09} with an associated uncertainty, it is estimated to be 0.48 - 1.1 \ms pc$^{-2}$ Gyr$^{-1}$,  which covers the results obtained from other studies including \citet{Rocha-Pinto_00}. On the other hand, the predicted SFR is 0.82  \ms pc$^{-2}$ Gyr$^{-1}$. We see a good agreement with the observations.

\begin{figure*}
\includegraphics[width=320pt]{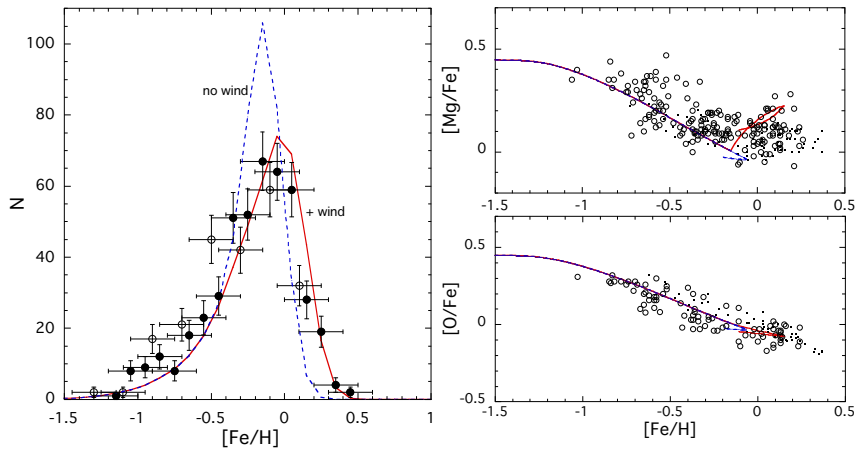}
\caption{Features of chemical evolution in the solar neighborhood. {\it left panel}: Predicted abundance distribution function of thin disk stars against the Fe abundance. A good fit to the observation is realized by the model with winds (solid line) while the result without winds is shown by the dotted line. The calculated distributions are convolved using the Gaussian with a dispersion of 0.05 dex as this is the minimum measurement error expected in the data. Open circles and crosses represent data taken from \citet{Edvardsson_93} and \citet{Wyse_95}, respectively.  The model distribution and the observed one by \citet{Wyse_95} are normalized to coincide with the total number of the sample stars used by \citet{Edvardsson_93}.{\it right panel}: Correlation of [Mg/Fe] and [O/Fe] with [Fe/H] for thin disk stars with our wind model and no wind model superimposed. The open circles and dots are taken from \citet{Edvardsson_93} and \citet{Bensby_05}, respectively.}
\end{figure*}

\subsection{gas fraction}

Finally, we see the evolution of gas fraction $f_g$ as a result of adopted infall rate and SFR since $f_g$ is a determinant factor to control the D/H abundance (the lower panel of Fig.~1). The predicted $f_g$ is compared with the observed value at the present time. An error in the observed $f_g$ mainly originates from a large uncertainty in the surface density of the ISM. Here we adopt its range from 6.5 \ms pc$^{-2}$\citep{Dame_93}  to 13  \ms pc$^{-2}$\citep{Kulkarni_87}, together with the stellar density of 35.5 \ms pc$^{-2}$. A gradual increase in $f_g$ over the last 4 Gyr while the SFR decreases seems at odds with the Schmidt-Kennicut law. However, a span in the SFR against a small range of $f_g$ in our result is compatible with the intrinsic scatter in the observed relation between the SFR and the surface density of the ISM \citep{Kennicutt_98}. What primarily induces this implied low star formation efficiency (SFE) irrespective of the steady supply of H I gas remains unclear, including the possible association with the low SFE deviated from the standard law in a low gas density environment such as 5-10 \ms pc$^{-2}$ observed for faint dwarf galaxies \citep{Roychowdhury_09} and low surface brightness (LSB) galaxies \citep{Wyder_09}. Interestingly, LSB galaxies exhibit about one-order change in SFR while those have similar gas densities. In addition, the Schmidt-Kennicutt law is valid only in a large scale such as $>$ 1 kpc 
\citep{Onodera_10}, whereas the local SFR observationally covers tens of pc.

\section{Heavy-Element Features}

We proceed to the next issue on chemical features of the local disk. Recall that we need to discuss the D/H evolution in the framework that gives a fully consistent explanation to the observed quantities on heavy elements such as Fe and $\alpha$-elements. The left panel of Figure 2 shows the successful reproduction of the ADF by our model. A stall of chemical enrichment owing to the decreasing SFR over the last Gyrs is compensated by the effective wind enrichment. In other words, the activity of local star formation is not sufficient to provide all of Fe contained in the existing long-lived stars. This is shown by the model case without the winds (dashed-line), which is skewed to the lower-metallicty, compared with the observations. 

This wind enrichment imprints its unique record on the elemental ratios of metal-rich stars. The winds are expected to exhibit the high [Mg/Fe] and low [O/Fe] ratios as mentioned in the model description. Accordingly, as a result of their reflections, our model predicts an upturning feature of [Mg/Fe] for metal-rich stars through a rapid increase-gradual decrease loop in the [Mg/Fe]-[Fe/H] correlation, which is in  good agreement with the observed data (the right upper panel). The different [O/Fe] evolution can be considered in the same theoretical scheme with an assumption of a low [O/Fe] in the winds (the right lower panel).

\begin{figure}
\includegraphics[width=230pt]{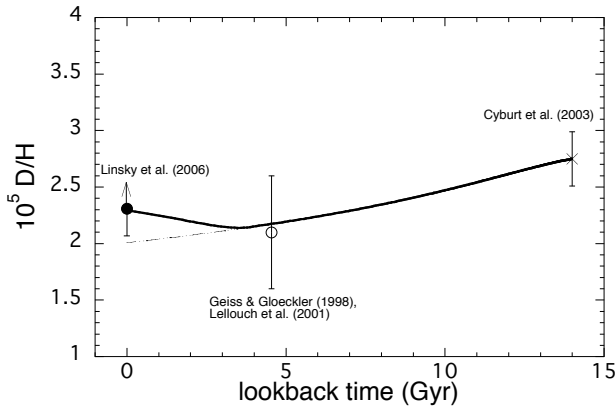}
\caption{The evolution of D/H abundance in the local disk predicted by the wind+decreasing SFR model (solid line). For reference, the model case with a constant SFR coefficient is shown by the dashed line. Three observed points are for the primordial value \citep{Cyburt_03}, the value for the protosolar cloud \citep{Geiss_98, Lellouch_01}, and the local ISM  value \citep{Linsky_06}.
}
\end{figure}

\section{Evolution of D}

Now we are all set to start on the discussion of D/H evolution. The predicted time-evolution of D/H is shown in Figure 3. Here the primordial (D/H)$_p$ is set to be 2.75$\times 10^{-5}$ \citep{Cyburt_03}. From its initial value, the D/H abundance starts to decrease with time, and subsequently its change turns to a gradual increase 4 Gyr ago. This turnover results from the supply of primordial D by an infall that overwhelms the destruction of D through star formation owing to its low rate. Then, it finally enables to achieve a present-day high D/H abundance of 2.3$\times 10^{-5}$, fully consistent with the observed value \citep{Linsky_06}, after passing D/H $\sim 2.1\times 10^{-5}$ equivalent to the protosolar value \citep{Geiss_98, Lellouch_01} around 4.6 Gyr ago. For reference, the case with a constant $\nu$ (=0.6) over the full age is indicated by dashed line.

Our safe conclusion is that the D/H abundance has little evolution during the last several Gyr, as if the D astration through star formation has turned off. It's because the metallicity of an infall would change in accordance with the cosmic evolution of damped Ly$\alpha$ systems \citep{Wolfe_05}, and eventually become subsolar with a smaller D than the primordial value like Complex C \citep{Sembach_04}. This view will lead to a flattening of D/H evolution rather than its gradual increase.

\section{Uncertainties in Observational Constraints}

In this section, we briefly review the uncertainties in two key observational constraints that largely influence the results presented here. As discussed below, neither of two has reached a firm consensus yet, thus further studies to better understand these issues, observationally and theoretically,  should be still awaited.

\subsection{recent star formation rate}
In our models, we adopt a sharply decreasing SFR which changes a factor of $\sim 4$ in the last 4 Gyr deduced by \citet{Fuchs_09}.  Their result is broadly consistent with other several works using different methods \citep{Hernandez_00, Vergely_02, de la Fuente Marcos_04, Cignoni_06}. However, at least two studies present a different picture for the recent SFR. \citet{Rocha-Pinto_00} derive the star formation history, using a chromospheric age distribution of dwarf stars, and their major finding is  that the disk has experienced a few bursts in the past, in which one of enhanced episodes of star formation occurred during the last 1 Gyr. On the other hand, from a new reduction of Hipparcos data combined with updated isochrones, \citet{Aumer_09} obtain the deceasing SFR from the past to the present but at a very small rate. 

With no doubt, major progress on this issue will come from a huge database brought by GAIA in the next decade.  

\subsection{present-day D/H abundance}

The controversial situation on the representative value of (D/H)$_0$ is nicely reviewed in \citet[][and references therein]{Tosi_10}. Although most claims seem inclined to be above (D/H)$_0$ $\sim 2\times 10^{-5}$, that is beyond the predictions by the conventional GCE models, very low values such as (D/H)$_0 \sim$1.5$\times 10^{-5}$ of the local bubble value or $\sim$0.98$\times 10^{-5}$ of the  highest column density regions \citep{Hebrard_05} can not be still discarded. The difficulty lies in the interpretation of a widespread variation in the observed gas-phase D/H ratios among local ISM regions. It seems reasonable to claim that the ISM exhibiting a high D/H such as $\sim$2.2-2.3$\times 10^{-5}$ is a result of incompletely mixed infall, i.e., a strong influence by a low-metal infall \citep[e.g.,][]{Steigman_07} since Complex C has a similar D/H of 2.2$\times 10^{-5}$. However, it is stemmed by a  counterargument that their O abundance is not so low as that expected in an infalling gas \citep{Hebrard_10}. 

Recently, analysis of Linsky's data by a Bayesian approach gives the best estimate of (D/H)$_0 \sim$2.0$\times 10^{-5}$ \citep{Prodanovic_10}. A reduction of $\sim 0.3\times 10^{-5}$ draws the observed (D/H)$_0$ close to the compatibility with some of GCE models. However, considering an uncertainty in the adopted probability distribution for depletion as well as an indication of high D in WNM \citep{Savage_07},  further studies with more data is still necessary for the settlement of a true (D/H)$_0$. Besides, recall that the observed D/H abundance only gives a lower limit of (D/H)$_0$.

To date, a high (D/H)$_0$ as proposed by \citet{Linsky_06} is viewed with a skeptical eye owing to an incompatibility with the theoretical GCE scheme. Here we propose a new path to end up with (D/H)$_0\sim$2.3$\times 10^{-5}$. Accordingly, we claim a need to have a fresh look at those observed high D/H, bearing in mind that they are within the predictions of a renewed GCE picture. 

\begin{figure}
\includegraphics[width=230pt]{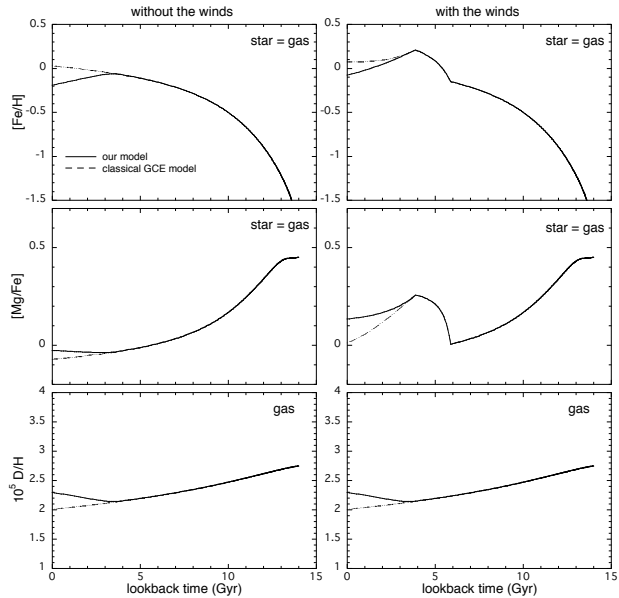}
\caption{Time evolution of [Fe/H], [Mg/Fe], and D/H. {\it left panels}: Solid and dashed lines represent the results by the model with the decreasing SFR and  with a constant SFR coefficient, respectively, under the assumption of no winds. {\it right panels}: Same as in left panels but for an introduction of the winds.
}
\end{figure}

\section{Roles of SFR and winds}

Finally, let's review each role of the sharply decreasing SFR and the metal-rich winds that are the main drivers of the successful results through the comparison of two models. Left panels of Figure 4 show the time evolution of [Fe/H], [Mg/Fe], and D/H calculated with the model in which the sharply decreasing SFR for the last 4 Gyr is incorporated into the GCE scheme. These witness that an inactive SF period makes the D/H evolution compatible with the observation, while leaving the problem of an incomplete chemical enrichment. Then, the satisfactory enrichment is achieved by an introduction of the metal-rich winds into the model with the D/H feature unchanged (right panels). For reference, in both models, the results calculated with a constant SFR coefficient are indicated by dashed lines.

\section{Conclusions}

As long as we stick to the conventional scheme of chemical evolution, recent observational implications of (i) the present-day high D/H abundance and (ii) little evolution of its abundance during the last several Gyr in the local disk are hard to believe. Accordingly, the discussion on it so far has been inclined to favor a much lower present-day D/H abundance in terms of an uncertainty in its estimate mainly due to dust correction. 

Here we present a new view on the D/H evolution in the local disk, based on the theoretical scheme that considers  the wind enrichment and the decreasing SFR with a rate of $\sim$ -0.8 \ms Gyr$^{-1}$, two phenomena that are implied to have occurred in the last several Gyr from the observed feature of local disk stars. The essence of its theory is that all heavy elements are not locally produced but some fraction comes form the outside, thus the models whose mission is the construction of the metal-rich system accept the observed low activity of star formation in the recent Gyrs, and end up with the ISM rich in D.

We insist that the local star formation does not suffice to produce all elements in-situ, and thus D does not destroy so much as expected by the chemical features of the local disk. Similar claim may be seen in the mechanism of radial mixing \citep{Roskar_08b, Schoenrich_09}, which brings metal-rich stars from the outside (i.e., the inner disk) instead of heavy elements. \citet{Schoenrich_09} show the star formation history for the local disk which exponentially decreases 1 Gyr after the beginning of  star formation, as implied by the new reduction of Hipparcos data \citep{Aumer_09}. Their model succeeds in a high-Z stellar abundance system as an outcome of radial migration of stars while it will surely predict a low D today. 

We strongly predict an ongoing infall onto the Galaxy disk. Our claim is that it operates to reduce the gas abundances in the disk. Indeed, today's observed gas abundances are not metal-rich at any places of the disk. The metallicities of Cepheids at the inner disk ($R\gtsim$ 4kpc) are [Fe/H]$\sim +0.2-0.3$ at the most \citep{Andrievsky_04}. Moreover, a surprisingly rsult is brought by supergiants in two young clusters in the innermost disk at $R$=3-4 kpc, whose metallicities are found to be [Fe/H]$\sim$ -0.1$-$ -0.2 \citep{Davies_09}. An infall must rain down on today's disk.

\section*{Acknowledgments}
Discussions with J. Bland-Hawthorn are gratefully acknowledged. The author wishes to thank the referee for all valuable comments that helped improve the paper, and is assisted in part by Grant-in-Aid for Scientific Research (21540246) of the Japanese Ministry of Education, Culture, Sports, Science and Technology.

\label{lastpage}

\end{document}